\documentclass[12,leqno]{article} 

\newcommand{\Red}{\color{black}}
\newcommand{\Blue}{\color{black}}

\newcommand{\m}[1]{ {\Red $#1$} }

\newcommand{\Em}{ \em\color{black}   }
\newcommand{\beq}{\Red \begin{eqnarray}}
\newcommand{\eeq}{\nonumber\end{eqnarray}\Blue}
\newcommand{\numeq}{\end{eqnarray}\Blue}
\newcommand{\half}{ {1\over 2} }

\newcommand{\eps}{\epsilon}

\usepackage{color}
\usepackage{graphics}

\newcommand{\beqs}{\begin{eqnarray}}
\newcommand{\eeqs}{\end{eqnarray}}
\def\sgn{\;{\rm sgn}\;}
\def\m#1{$#1$}

\def\Ai{\;{\rm Ai}\;}

\begin{document}

\centerline{\bf\large A Model of  Interacting Partons for Hadronic Structure
Functions}

\centerline{G. S.  Krishnaswami\footnote{govind@pas.rochester.edu} and
S. G. Rajeev\footnote{rajeev@pas.rochester.edu}}

\begin{center}
   {\it Department of Physics and Astronomy, University of Rochester, 
    Rochester, New York 14627} \\
   \vspace{.5cm}

   \vspace{2cm}
   {\large\bf Abstract}
\end{center}

We present a model for the structure of baryons in which the valence 
partons  interact  through  a linear potential. This  model can be
derived  from QCD in the approximation where transverse momenta are
ignored. We compare  the
baryon structure function predicted by our model with a standard
global fit to  Deep Inelastic
Scattering data. The only parameter we can adjust is the fraction of the
baryon momentum  carried by the valence partons.
Our prediction agrees  well with data except for small values of the 
Bjorken scaling variable.  



\pagebreak

The quark-parton model \cite{parton} was proposed as a  simple
explanation of  the phenomenon of scaling in Deep Inelastic
Scattering. After the discovery of Quantum Chromodynamics (QCD),  scaling
was understood  as a consequence of asymptotic freedom
\cite{af}. Moreover,
 QCD predicts small
violations of scaling that can be calculated perturbatively: knowing
the parton distribution at one (large enough) energy scale, it is
possible to  predict them  at any other  large energy. Although these
predictions have been confirmed impressively,  there is as yet no theoretical
understanding of the initial parton distributions; they are controlled
by non-perturbative effects. Accurate knowledge of the parton
distributions is essential  in order to make predictions
about any high energy scattering process in a hadron collider. Also,
only after we understand the parton distributions as predicted by the
standard model, can we look for deviations which may signal a
substructure for quarks. 
As a result 
 tremendous effort has been expended by several groups of theorists
and experimentalists to extract the parton distributions from
scattering data at various energy scales \cite{CTEQ,MRST,GRV}.

We will present a model of interacting partons for  hadronic
structure functions. 
The partons are assumed to be  relativistic
particles interacting with each other through a linear
potential, which is a sensible idea in the lightcone
formalism. Their momenta transverse to the direction of the collision
will be ignored. We will solve for the parton wavefunction within the
 approximation that it  factorizes into a product of
single particle wavefunctions. This is  analogous to the Hartree--Fock
approximation of atomic physics. After these approximations, the
wavefunction can be determined as the solution of a nonlinear integral
equation. This equation can be solved numerically, although we also
find a good  approximate solution by analytic methods in some special cases. 

We can compare with the structure functions extracted by several
groups of experimentalists  and theorists. Considering that we have at
most one  parameter to adjust, the agreement is quite good, except for small
values of the Bjorken scaling variable \m{x_B}. In this region our
model is not expected to be valid: sea quarks and gluons cannot be ignored.

Can we derive this interacting  parton model from  QCD? 
In fact we have already given such a derivation
in  previous publications \cite{twodhadron}. In the
approximation in which transverse momenta are ignored, two of the
space--time co-ordinates become irrelevant: QCD is replaced by its
dimensional reduction to two dimensions. That such a `collinear QCD'
can describe hadronic structure functions has also been proposed by
other scientists \cite{brodsky}. What is unique to our approach is the
picture of a baryon as a topological soliton, which remarkably enough,
gives a  derivation of  our interacting parton model from first principles.
In previous work,
 \cite{twodhadron} two dimensional QCD was 
shown in the large \m{N} limit to be equivalent to a  
bilocal field theory whose phase space is an infinite dimensional
Grassmannian. The baryon is a topological soliton in this theory,
whose energy and structure can be estimated (within a variational
approximation) by a nonlinear integral equation. This is precisely the
integral equation we will study here. After the equation was derived
and studied in this way it became clear that it had a simple interpretation in
terms of the parton model. It is this parton model point of view that
we will mostly use in this paper.

In this approximation, the dynamical degrees of freedom associated with
gluons are ignored: only the longitudinal components of the gluons
which can be eliminated in favor of the quarks are kept. In later
publications we will study how tranverse gluons will modify the
structure of a baryon; in particular derive the gluonic distribution 
functions.

The  stucture functions of a  hadron are measured in deep inelastic collisions
of a virtual photon ( or weak gauge boson) with a proton or neutron.
 Since the component of 
momentum  in the direction of collision is much greater than those
transverse, it is usually  assumed that all the constituents
of the hadron (partons) are moving in the same direction. We will
choose this direction to be the \m{x^1}-axis. It is also customary to
use the null component of momentum \m{p=p_0-p_1} instead of the
spatial component as the basic kinematic variable \cite{parton,CTEQ}. 
In terms of the energy \m{p_0} and null momentum, the mass shell
condition for  a particle of mass \m{m} is
\beq
	-p^2+2p_0p=m^2
\eeq
or equivalently, 
\beq
	p_0=\half[p+{m^2\over p}].
\eeq
Note that since \m{p_0=\surd(p_1^2+m^2)>p_1}, the null component of 
momentum \m{p=p_0-p_1} is always positive. 
\footnote{The mass shell condition looks a bit unusual beacuse 
 the metric tensor is not
diagonal in our choice of variables. If we  use the conjugate
variables \m{(u,x)} to \m{p_0,p}, $ds^2=du^2+2dudx$. Here \m{x} is the
usual spatial co-ordinate and \m{u=x^0-x} is a null co-ordinate. }

Let \m{\tilde\psi(\nu,\alpha, p)} be the wavefunction of a single parton
 expressed as a
function of the null component of momentum. Here \m{\alpha=1,\cdots M}
is a discrete quantum number that labels spin and flavor; and \m{\nu=1,\cdots
 N} labels color. Although the  value of color in nature is three, we
 will find it convenient to keep it arbitrary for now.
 Because of the positivity
condition on \m{p},
\beq
	\tilde\psi(\nu,\alpha,p)=0,\; {\rm for}\; p<0.
\eeq
A baryon, which  is made up of \m{N} such partons, will have a
 wavefunction
 \m{\tilde\psi(\nu_1,\alpha_1,p_1;\nu_2,\alpha_2,p_2;\cdots;
 \nu_N,\alpha_N,p_N)}. This will vanish unless all the momenta are
 positive. Since the partons are fermions (after all, they are quarks), the
 wavefunction must be totally anti--symmetric under permutations of  the
 partons. However, the baryon must be colorless  so the 
 color quantum numbers must be completely antisymmetric by themselves:
\beq
\tilde\psi(\nu_1,\alpha_1,p_1;\nu_2,\alpha_2,p_2;\cdots,
 \nu_N,\alpha_N,p_N)=\eps_{\nu_1,\nu_2,\cdots \nu_N}
\tilde\psi(\alpha_1,p_1;\alpha_2,p_2;\cdots \alpha_N,p_N)
\eeq
where \m{\eps_{\nu_1,\nu_2,\cdots \nu_N}} is the Levi-Civita tensor.
So we can forget about color and deal with the remaining part of the
 wavefunction \m{\tilde\psi(\alpha_1,p_1;\alpha_2,p_2;\cdots;
 \alpha_N,p_N)} which is completely {\Em symmetric}.

 The kinetic energy of a system of
\m{N} partons will be, 
\beq
\sum_{\alpha_1\cdots \alpha_N}\int_0^\infty \sum_{i=1}^{N}\half
[p_i+{m_{\alpha_i}^2\over
p_i}]|\tilde\psi(\alpha_1,p_1;\cdots; \alpha_N,p_N)|^2{dp_1\cdots dp_N\over (2\pi)^N}.
\eeq

If there is a two body  potential \m{g^2v(x)}between the partons,the total
potential energy will be \footnote{ A  multiplicative factor has been
chosen such that  the coupling constant \m{g} has units of mass.}
\beq
\half g^2\sum_{\alpha_1\cdots \alpha_N}\int_0^\infty 
\sum_{i\neq j }
v(x_i-x_j)|\psi(\alpha_1,x_1;\cdots; \alpha_N,x_N)|^2 dx_1\cdots
dx_N
\eeq

Here,
\beq
	\psi(\alpha_1,x_1;\cdots \alpha_1,x_N)=\int_0^\infty \tilde\psi(\alpha_1,p_1;\cdots\alpha_N, p_N)
 e^{i\sum_j p_jx_j}{dp_1\cdots dp_N\over (2\pi)^N} 
\eeq
is the wavefunction in position space.

In addition to the kinetic and potential energies, there could be a
term in the hamiltonian describing the self--energy  of the partons. 
Due to Lorentz
invariance, such a term can only be  a (finite) renormalization of
\m{m^2}; i.e., it is of the form 
\beq
c\sum_{\alpha_1\cdots \alpha_N}\int_0^\infty \sum_{i=1}^{N}
{g^2\over
2\pi p_i}|\tilde\psi(\alpha_1,p_1;\cdots; \alpha_N,p_N)|^2{dp_1\cdots dp_N\over (2\pi)^N}
\eeq
for some constant \m{c}.

 Thus, the ground state
wavefunction is determined by minimizing the total energy 
\beqs
	{\cal E}_N(\tilde\psi)&=&\sum_{\alpha_1\cdots \alpha_N}\int_0^\infty \sum_{i=1}^{N}\half[p_i+{\mu_{\alpha_i}^2\over
p_i}]|\tilde\psi(\alpha_1,p_1;\cdots \alpha_N,p_N)|^2{dp_1\cdots
dp_N\over (2\pi)^N}\cr
 & &+ \half g^2 \sum_{\alpha_1\cdots \alpha_N}\int_0^\infty
\sum_{i\neq j}
v(x_i-x_j)|\psi(\alpha_1,x_1;\cdots \alpha_N,x_N)|^2 dx_1\cdots
dx_N.\nonumber
\eeqs
Here \m{\mu_\alpha^2=m^2+c{g^2\over \pi}} is the renormalized mass of
the parton. 

The success of the simple parton model and its QCD inspired variants
shows that it is a good approximation to describe the system in terms
of  particle distributions that depend only on the momentum of a single
parton. In other words  correlations between the
partons can be ignored, except in that the total momentum of the
partons must be fixed. 
Thus it  should be a good approximation to assume that (in the ground state),
the baryon  wavefunction has the form  :
\beq
	\tilde\psi(\alpha_1,p_1;\cdots \alpha_N,p_N)=
2\pi\delta(P-\sum_ip_i)\prod_{i=1}^N\tilde\psi(\alpha_i,p_i).
\eeq
Here, \m{P} is the momentum of the baryon. Since each of the parton
momenta \m{p_i} are positive, we see that 
\beq
	\tilde\psi(p)=0\;{\rm unless}\; 0\leq p\leq P.
\eeq

This ansatz is  analogous to  the Hartree approximation of atomic physics,
which works well even when the number of electrons is small as in the
Helium atom. In any case we can regard our product  as a variational ansatz
for the ground state of the baryon. We will normalize the single
parton wavefunction to have length one:
\beq
	||\tilde\psi||^2=\sum_{\alpha=1}^M\int_0^P|\tilde\psi(\alpha,p)|^2 {dp\over 2\pi}=1.
\eeq
Moreover it must satisfy the   sum rule on momentum: the  total
momentum of all the partons must equal  the momentum of the baryon:
\beq
	N\int_0^P p|\tilde\psi(p)|^2{dp\over 2\pi}=P.
\eeq

In fact the valence partons may  not carry all the momentum of the
baryon. (The structure functions extracted from data show that only
about half the momentum of the baryon is carried by the valence
partons.) We can allow for this by replacing the momentum sum rule
above by 
\beq
	N\int_0^P p|\tilde\psi(p)|^2{dp\over 2\pi}=fP
\eeq
where \m{f} is the fraction of the baryon momentum carried by all the
valence partons. We will see that this fraction \m{f} is the only
parameter on which the sturcture functions will depend.

The energy {\Em per parton}  becomes now
\beqs
	E&=&\sum_\alpha\int_0^P \half[p+{\mu_\alpha^2\over
p}]|\tilde\psi(\alpha,p)|^2{dp\over 2\pi}+\cr
  & & \half{\tilde{g}^2}\int_{-\infty}^\infty v(x-y)
\sum_\alpha|\psi(\alpha,x)|^2\sum_\beta|\psi(\beta,y)|^2dxdy.
\eeqs
where, \m{\tilde g^2=g^2N} and
\m{\psi(\alpha,x)=\int_0^P\tilde\psi(p)e^{ipx}{dp\over 2\pi}}. 

What potential \m{v(x)} should  we use? It is known that the
potential between quarks is, to a good approximation, linear. Moreover
the collinear approximation to QCD mentioned above predicts a linear
potential. Hence we will choose 
\beq
	v(x)=\half |x|.
\eeq
A linearly increasing  potential can lead to an infrared divergence in
the energy  of  a wavefunction, even  one  that is  decaying at
infinity. For example if \m{\psi(\alpha, x)\sim {1\over x} } for large \m{x}
the potential energy term will diverge logarithmically. This is where
a proper choice of the self-energy term mentioned above comes in. If
we choose the constant \m{c} in the self-energy such that 
\beq
	\mu_\alpha^2=m_\alpha^2-{\tilde g^2\over \pi},
\eeq
this Infrared divergence dissappears. This is indeed the value of the
self-energy term predicted by ( the large \m{N} limit of ) two
dimensional QCD.

The wavefunction of the partons in a baryon is now determined by
minimizing the above energy function subject to the constraints on the
wavefunction
\beq
	\tilde\psi(p)=0\;{\rm for }\; p<0\;{\rm and for }\; p>P\;{\rm and}\;
||\tilde\psi||^2=1.
\eeq

The   minimization of  the above energy is equivalent to the solution
of a system of nonlinear integral equations. 
The condition that the wavefunction
vanish outside of the interval \m{[0,P]} in {\it momentum} space
 is a subtle constraint on the {\it
position} space
wavefunction: \m{\psi(\alpha, x)} must be the boundary value of an
entire function which grows at most like \m{e^{P|x|}} at infinity. 
Since such a condition is
clearly difficult to impose numerically, we work in momentum space.
Variation of the energy with respect to \m{\tilde \psi} gives
\beq
\bigg[\half(p+{\mu_\alpha^2\over p})-\lambda\bigg]\tilde\psi(\alpha,p)+{\tilde
g^2}{\cal P}\int_0^P\tilde{V}(p-q)\tilde\psi(\alpha,q){dq\over 2\pi}=0
\eeq
where
\beq
	\tilde{V}(p)=-{1\over p^2}\sum_{\alpha}\int_0^P
\tilde\psi^*(\alpha,p+q)\tilde\psi(\alpha,q){dq\over 2\pi}.
\eeq
Here, \m{\lambda} is the Lagrange multiplier that enforces the
constraint \m{||\tilde \psi||^2=1}. (Since the function we are
minimizing is quartic in \m{\tilde\psi} this is {\Em not} the same as
the  minimum energy \m{E}.)

The second equation is just the momentum space version of the equation 
for the mean potential due to all the partons:
\beq
	V''(x)=|\psi(x)|^2.
\eeq
Note that the Fourier transform \m{\tilde{V}(p)=\int_{-\infty}^\infty
V(x)e^{-ipx}dx} of the mean potential is singular at the origin:
\beq
	\tilde{V}(p)\sim-{1\over p^2}\;{\rm for}\; p\to 0.
\eeq
Hence the integrand  in the  equation for \m{\tilde\psi(\alpha,p)} is
singular. We must define it to be a finite part integral in the sense
of Hadamard \cite{hackbusch}:
\beq
	{\cal P}\int_{-\infty}^\infty \tilde{V}(p-q)\tilde\psi(\alpha,q){dq\over 2\pi}
=\int_{0}^\infty[\tilde \psi(\alpha,p+q)+\tilde\psi(\alpha,p-q)-
2\tilde\psi(\alpha,p)]\tilde V(q){dq\over 2\pi}.
\eeq

 By studying the
behavior of the equation in the neighborhood of the singular point
\m{p=0}, with the ansatz \m{\tilde\psi(p)\sim p^{\nu}}, we get
the following formula relating the exponent \m{\nu} and \m{m^2}:
\beq
	{\pi m^2\over \tilde g^2}=1+\int_0^1{dy\over y^2}\bigg[
 (1+y)^{\nu}+ (1-y)^{\nu}-2\bigg]+
\int_1^\infty {dy\over y^2}\bigg[(1+y)^\nu-2\bigg].
\eeq
Thus, at the critical point \m{m=0}, we have \m{\nu=0}: the critical 
 wavefunction
tends to a constant as \m{p\to 0}. If \m{m^2>0}, the wavefunction
vanishes like a power: \m{\nu>0}.

This nonlinear integral equation was derived previously 
\cite{twodhadron} from the large
\m{N}-limit of two dimensional QCD. The  `master field' of this limit
of 2DQCD is a hermitean matrix \m{\tilde
M(\alpha,p;\beta,q)} satisfying the nonlinear constraint
\beq
	\tilde M(\alpha,p;\beta,q)[\sgn(p)+\sgn(q)]+\sum_{\gamma}\int {dr\over 2\pi}\tilde
M(\alpha,p;\gamma,r)\tilde M(\gamma,r;\beta,q)=0.
\eeq
The set of solutions of this condition is an infinite dimensional
Grassmannian manifold, which has connected components labelled by an
integer: the `virtual rank', \m{-\half\sum_\alpha\int \tilde M(\alpha,p;\alpha,p){dp\over
2\pi}}. This integer has the physical 
meaning of baryon number. 
The baryon is thus a topological soliton and is described by a
configuration of minimal energy in the sector of virtual rank one.

Thus we have here a realization of Skyrme's idea that the baryon is a
topological soliton\cite{skyrme}.
The energy of the configuration \m{\tilde M(p,q)} is given by 
\beq
	{\cal E}(\tilde M)=\sum_{\alpha}\int \half\big(p+{\mu_\alpha^2\over p}\big)\tilde 
M(\alpha,p;\alpha,p){dp\over
2\pi}
	+{\tilde g^2}\sum_{\alpha,\beta}\int |M(\alpha,x;\beta,y)|^2v(x-y)dxdy.
\eeq
Here,
\beq
	M(\alpha,x;\beta,y)=\int \tilde M(\alpha,p;\beta,q)
e^{ipx-iqy}{dpdq\over (2\pi)^2}.
\eeq
The parameters \m{\mu^2_\alpha} and \m{\tilde g} are related to those of 2DQCD:
\beq
	\mu_\alpha^2=m_\alpha^2-{\tilde g^2\over \pi},\quad \tilde g^2=g^2N
\eeq
where \m{m} is the current quark mass and \m{g} the gauge coupling
constant. Note that in the limit of chiral symmetry, \m{m_\alpha=0}, (when the lightest
meson is massless) and the value of \m{\mu^2_\alpha} is actually negative:
\beq
	\mu_\alpha^2=-{\tilde g^2\over \pi}.
\eeq
This is the value we will mostly study.

In addition to topological soliton solutions describing the baryon the
above equation also describes small fluctuations from the vacuum. The
corresponding particles are the mesons. If we assume that
\m{\tilde{M}(p,q)} is infinitesimally small, its equations of motion
will reduce to a {\it linear } integral equation. This equation was
first derived by 't Hooft by a masterful use of diagrammatic methods
\cite{thooft}. Witten \cite{witten} suggested later that the baryon can be described
by a Hartree-Fock aproximation in the large \m{N} limit of QCD. He
carried out this idea in a non-relativistic context. Our
model may be thought of a relativistic implementation of this idea of Witten.

Now if we make the variational ansatz that 
\m{\tilde M(\alpha, p;\beta,q)=-2\tilde \psi(\alpha,p)\tilde \psi^*(\beta,q)} is
separable, the quadratic constraint on \m{\tilde M} becomes the
condition that 
\beq
	\tilde\psi(\alpha,p)=0\;{\rm for}\; p<0.
\eeq
Moreover, the energy of the soliton reduces to just the formula we
obtained within the parton model! (The momentum sum rule is the 
the analogue of the WKB quantization condition in this theory.)
This is how we can derive the
parton model from collinear QCD. It is remarkable that the topological
soliton model, which at first looks so different from the parton model, leads
to exactly the same equation for the hadron structure function. 

The variational approximation that \m{\tilde M} be separable
corresponds in the parton model to ignoring the `sea quarks'. This
seems reasonable only for values of the Bjorken parameter \m{x_B} that
are not too small. In general we should get better answers by
directly minimizing the energy of the soliton model without the
further  approximation of the separable ansatz. We are currently
studying this issue.

Let us return to solving the integral equation.
It cannot be solved
analytically, so we will resort to  a numerical method. ( We can get approximate analytical solutions in some
special cases which agree well with the numerical solution.) 
For simplicity, 
we will from now on put all the parton masses
equal to  each other. It is not difficult to modify our solutions to
include unequal masses. Also, it is sufficient to  find a solution that is non--zero
only for one value of the spin-flavor index \m{\alpha}:
\beq
	\tilde\psi(\alpha,p)=\delta_{\alpha,1}\tilde\psi(p).
\eeq
This breaks the \m{U(M)} invariance of our model spontaneously. This
symmetry can   be restored later by the collective variable method as in
the theory of solitons, but we will not address this issue here. 

Let us first describe our approximate analytic solution for the
special case \m{\mu=0} and large \m{N}. We are mostly interested in the case
\m{\mu^2=-{\tilde g^2\over \pi}} but this approximate solution will
help to validate our numerical procedure.
It is reasonable to expect that most of the
contribution to the integral 
\beq
\int_{0}^\infty[\tilde\psi(p+q)+\tilde\psi(p-q)-2\tilde\psi(p)]\tilde{V}(q){dq\over 2\pi}
\eeq
will come from the neighborhood of the singularity near \m{q=0}:
\beq
\int_{0}^\infty[\tilde\psi(p+q)+\tilde\psi(p-q)-2\tilde\psi(p)]\tilde V(q){dq\over
2\pi}\sim \tilde\psi''(p)\int_0^\infty q^2\tilde V(q) {dq\over 2\pi}.
\eeq

Set  
\beq
	a=-\int_0^\infty q^2\tilde{V}(q) {dq\over
2\pi}.
\eeq
Using the fact that the ground state wavefnction is real, we can show
that 
\beq
	a=\half |\int_0^\infty\tilde\psi(p){dp\over 2\pi}|^2.
\eeq

Thus  we get a differential equation:
\beq
	-\tilde{g}^2 a \tilde\psi''(p)+\half[p+{\mu^2\over
p}-2\lambda]\tilde\psi(p)=0.
\eeq
This can be thought of as a Schr\"odinger equation with a linear
potential  plus a Coulomb potential; such an equation  has been used to study
heavy quark-anti-quark bound states \cite{quarkonium}. But the
physical origin of our  equation is completely different:  we get
this equation in momentum space {\Em not} position space. Also, our
quarks are not assumed to be  heavy and  we are studying  a baryon not
a meson.

In the special case \m{\mu^2=0},  the solution is a linear combination
of Airy functions.
 But it is awkward to work with this linear
combination. If we make the further approximation that \m{N} is
large, our momentum sum rule shows that \m{P} is large as
well. Then the boundary condition on the wavefunction is that it must
tend to zero at infinity: in other words 
\beq
	\tilde\psi(p)=
C \Ai\big({p-2\lambda\over (2a\tilde{g}^2)^{1\over 3}}\big)
\eeq

The eigenvalue \m{\lambda} is determined by the
condition that the solution vanish at the origin:
\beq
	\lambda=-\half\xi_1(2a\tilde{g}^2)^{1\over 3}.
\eeq
Here, \m{\xi_1\sim -2.33811} is the root of the Airy function closest
to the origin. \m{C} is determined by the normalization condition.

The constant \m{a} is determined by putting this solution back into
the definition of \m{a}. We get,
\beq
	2a=\tilde{g}{1\over (2\pi)^{3\over 2}}{|\int_{\xi_1}^\infty\Ai(\xi)d\xi|^3\over
\big[\int_{\xi_1}^\infty \Ai^2(\xi)d\xi\big]^{3\over 2}},\quad
	\lambda={|\xi_1|\over 2\surd(2\pi)}{\int_{\xi_1}^\infty
\Ai(\xi)d\xi\over \big[\int_{\xi_1}^\infty
\Ai^2(\xi)d\xi\big]^{1\over 2}}\tilde{g}\sim 0.847589 \tilde{g}.
\eeq
Moreover
\beq
	\tilde\psi(p)=C\Ai\big(\xi_1({p\over \tilde{g}\lambda}-1)\big).
\eeq

This  analytic approximation agrees well with the numerical
solution (described below) for \m{\mu^2=0} and large \m{N}. 
This confirms  the validity of our numerical method.

	Now let us discuss the numerical solution of this
problem. This is {\Em not}  straightforward since the kernel of the integral
equation is singular.  We need a reliable method of 
numerical quadrature for  integrals such as 
\beq
	{\cal P}\int_0^P f(p,q)\rho(q)dq
\eeq
when the weight function \m{\rho(q)} has a singularity at \m{q=0} like
\m{1\over q^2}. 
We
need to subdivide the interval
\m{[0,P]=\cup_{r=1}^{n}[b_r,b_{r+1}]} into subintervals. Within
each subinterval we choose a set of points \m{q_{jr}, j=1,\cdots
\nu_r}. We approximate the integral by a sum 
\beq
	\int_0^P f(p,q)\rho(q)dq=\sum_{jr}w_{jr}f(q_{jr}).
\eeq	
The weights \m{w_{jr}} are determined by the condition that within
each subinterval \m{[b_r,b_{r+1}]}, the integral of a polynomial of
order \m{\nu_r-1} is reproduced exactly:
\beq
 {\cal P}\int_{b_r}^{b_{r+1}} q^k
\rho(q)dq=\sum_{j=1}^{\nu_r}w_{jr}q^k_{jr}.
\eeq
This is just the usual method of numerical quadrature except that the
integral on the left hand side is singular for \m{b_r=0} and
\m{k=0,1}. In these cases we can evaluate the l.h.s analytically as
the finite part in the sense of  Hadamard. 
(The main difference from
the usual situation is that  the moments
on the l.h.s of the above equation are not all positive.) The weights are then
determined by solving the above system of linear equations.

Given an approximate mean potential \m{\tilde{V_s}(p)} we can convert
the linear integral equation 
\beq
	\half[p+{\mu^2\over p}-2\lambda_s]\tilde\psi_{s+1}(p)+{\tilde
g^2}{\cal P}\int_0^\infty\tilde{V_s}(p-q)\tilde\psi_{s+1}(q){dq\over 2\pi}=0
\eeq
into a matrix eigenvalue problem by the above method of
quadrature. We use the ground state eigenfunction so determined to
calculate numerically the next approximation \m{\tilde{V}_{s+1}} for  the mean
potential. This process is  iterated until the solution converges.
Having determined the wavefunction, we must impose the momentum sum
rule to determine \m{\tilde g}. 

 We chose \m{n=50} intervals each containing
\m{\nu_r=6} points for our quadrature. The resulting matrix was diagonalized
using Mathematica. As a starting mean potential we used
\beq
	\tilde{V}_0(q)=-{e^{-\big({q\over \tilde g}\big)^{3\over 2}}\over q^2}.
\eeq
The procedure converges in about \m{10} iterations. 
We find that due to the errors in the discretization of the problem,
the critical value of \m{\mu^2} (the value at which the wavefunction
goes to a  constant at \m{p=0}) is a bit lower (\m{-1.3{\tilde
g^2\over Pi}}) than the theoretical
value of \m{-{\tilde g^2\over \pi}}.

Now we turn to the question of the comparison of our model with data from
Deep Inelastic Scattering.  It is
customary to describe the parton distributions as a function of the
Bjorken scaling variable \m{0\leq x_B\leq 1} which is the fraction of 
the null component of momentum
carried by each parton. This means we will rescale momenta to the
dimensionless variable \m{p=x_BP}. The probability density of a parton
carrying a fraction \m{x_B} of the momentum is then 
\beq
	\phi(x_B)= {P\over 2\pi}|\tilde\psi(x_B P)|^2.
\eeq
(The factor of \m{1\over 2\pi} is needed because \m{\phi(x_B)} is
traditionally normalized to one with the measure \m{dx_B} rather than
\m{dx_B\over 2\pi}.)

It is important to note that the only dimensional parameter in our theory,
\m{\tilde g}, cancels out of the formula for \m{\phi(x_B)}: it only
serves to set the scale of momentum and when  the wavefunction is
expressed in terms of the  dimensionless variable
\m{x_B} it cancels out. We have set \m{\mu^2} to the critical value
(within numerical errors),
which is the value corresponding to chiral symmetry; i.e., zero current
quark mass. The number of colors we fix at \m{N=3}. 
Thus the only free parameter in our theory is  the fraction \m{f} of the baryon
momentum carried by the valence partons. The parameters \m{N} and
\m{f} appear in the combination \m{N_{\rm eff}={N\over f}}.

We have ignored 
the  ispospin  of the quarks in the above discussion. 
We should therefore compare our structure functions  with the isoscalar
combination of the valence quark distributions of a
baryon, \m{\phi(x_B)}. 
It is not difficult to take into account of isospin effects.

The parton distributions have been extracted from scattering data by
several groups of physicists \cite{CTEQ,MRST,GRV}.
In the accompanying figures we plot our wavefunctions for a few values of
\m{f}  to those extracted from data by the MRST collaboration. We
agree remarkably well with experiment  except for small values of
\m{x_B}. The agreement is best when the fraction of the baryon momentum carried by
the valence partons is about \m{f=0.6}.

  Our model does not predict
the observed behavior of the parton  distributions for small \m{x_B}:
our probability distribution tends to a constant for small  \m{x_B}
although due to numerical errors that is not evident in the plots. 
The
observed distributions have an integrable singularity there: roughly
speaking, \m{\phi(x_B)\sim x_B^{-0.5}} for small \m{x_B}.The approximations we made clearly break down in the small \m{x_B}
region: sea quarks can no longer be ignored, indeed even gluons need to
be considered. We will study these effects in future publications.

\centerline{\bf Acknowledgement}
This research was supported by the Department of Energy through the
grant DE-FG02-91ER40685  and by the Summer Reach program of the
University of  Rochester.

\pagebreak[4]

\centerline{\includegraphics{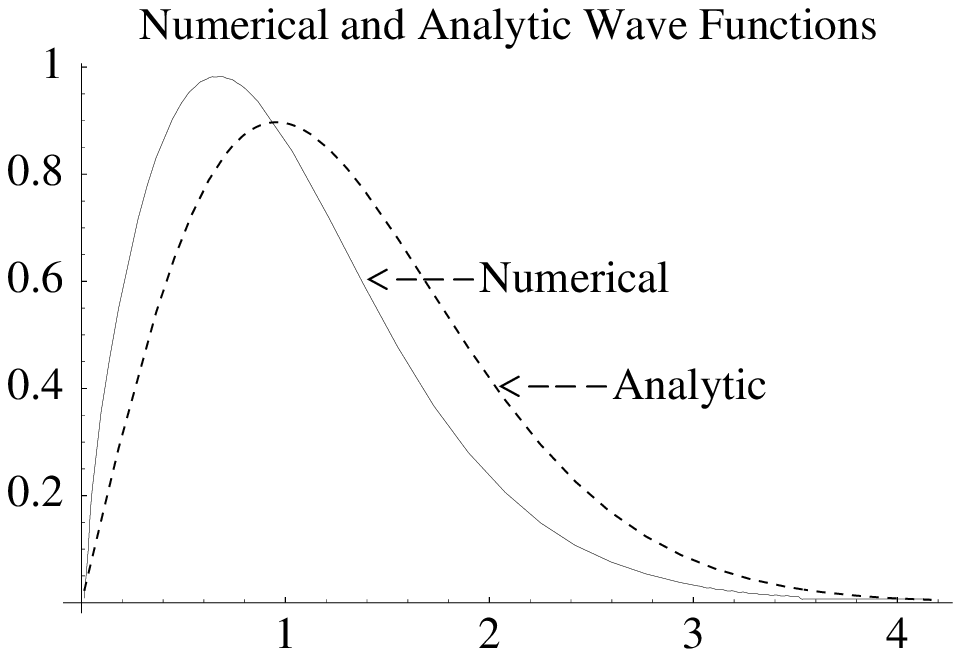}}

\begin{center}
{The numerical solution for \m{\mu^2=0,N=5} compared against the approximate
analytical solution. The horizontal axis is \m{p\over \tilde g} }
\end{center}

\centerline{\includegraphics{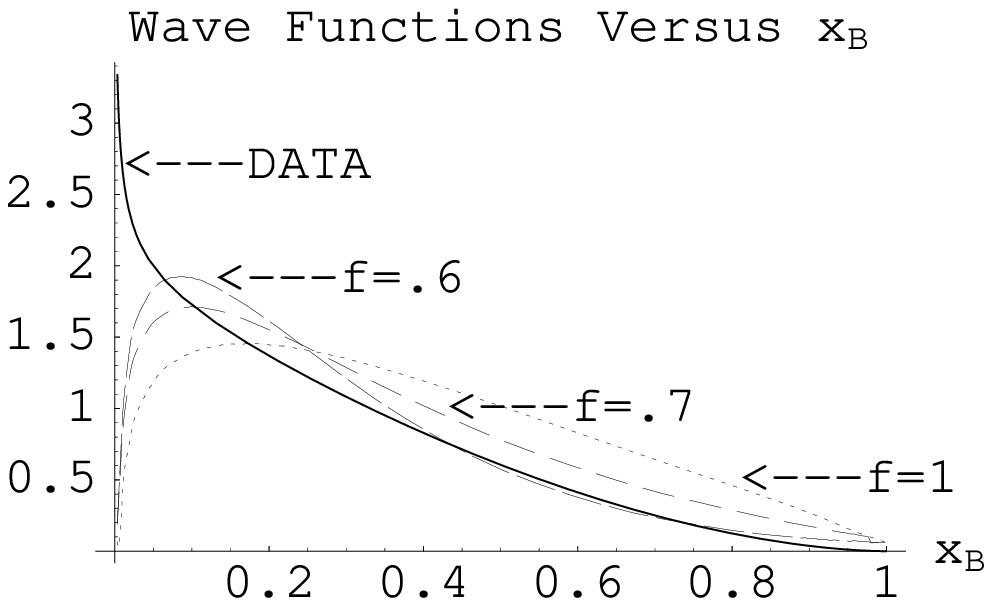}}

\begin{center}
{Comparison of  parton wavefunctions \m{\surd\phi(x)} with the MRST
global fit to data.}
\end{center}

\pagebreak[4]

\end{document}